# Self-Organization in Spontaneous Movements of Neonates generates Self-specifying Sensory Experiences

*Birte Assmann[123], Theresa Kaese[2], Anke Neumeister[1], Catherine Disselhorst-Klug[4]*

[1] *Freie Universität Berlin, Institute of Human Biology, Germany*
[2] *Niedersächsisches Institut für frühkindliche Bildung und Entwicklung, Germany*
[3] *Steinbeis Transfer Institut Gesundheit Bindung Bildung, Germany*
[4] *RWTH Aachen University, Institute of Applied Medical Engineering, Germany*

## Abstract

Movement experience and the coordination of perception and action are the basis of developing body awareness, emotion, motivation and cognition and the sense of self. The four limbs play a key role in the developing sense of body ownership, agency and peripersonal space.

Neonatal limb movements were investigated respective processes of self-organization and developing body awareness. Application of symbolic dynamics to kinematic data of radial distance trajectories of the hands and feet determined 16 coordination patterns according to the distance of the effectors to the body center. From time series analysis of occurrence and recurrence of these configurations of 25 movement episodes of 14 infants both characteristics of self-organization as well as features assumed to provide neonates with experiences involved in the development of body awareness were found. With increasing age a shift from configurations with proximal to distal positions suggests a role of the proximal-distal dimension in movement development.

We conclude that self-organization in spontaneous movements provides neonates with perceptual body- and self-specifying stimuli involved in developing body awareness and postulate the involvement of emotional and cognitive processes.

Keywords: neonates; spontaneous movements; kinematics; symbolic dynamics; self-organization; body awareness.

## 1. Introduction

Movement experience and the coordination of perception and action are the basis of the awareness of the body and the self as well as the interaction with the environment. Motor development includes experiencing and experimenting with possibilities and constraints that come along with corresponding feelings of e.g. interest, joy and anger. The pre- and early postnatal time is profound in laying major cornerstones of developing body awareness, emotion, motivation, cognition and the sense of self (Berlucchi & Agliot, 2010, Rochat 2012). The four limbs are our most extending and flexible body parts, define our peripersonal space and play a key role in the sense of body ownership and agency, which are basic constituents of the sense of self (Ehrsson, Holmes & Passingham, 2005; Kalckert & Ehrsson, 2012).

First movements of the four limbs are spontaneous movements of the fetus that can be observed at about the seventh week of gestagation. Muscle contractions are caused by excitation of nerve cells that contact the muscle cells. By sensory feedback, the corresponding synaptic connections of the developing nervous system are stabilized. Continuously from pre- to postnatal development neuronal representations emerge and form the basis of body awareness.

Limb trajectories of spontaneous movements are determined by their biomechanical and dynamic properties, such as interaction, centripetal and inertial forces, stiffness and viscoelasticity. It is triggered by spontaneous neuronal network activity and reflex pathways in the spinal cord. The latter are characterized by reflex irradiation and reciprocal coactivation in neonates (Myklebust & Gottlieb, 1993, Thelen & Fisher, 1983). Different research areas suggest an individually flexible development by self-organizing processes rather than an innate fixed wiring (Marques, Imtiaz, Iida & Pfeifer, 2013; Thelen & Fisher, 1982).

Spinal spontaneous network activity can autonomously generate spontaneous motor activity before and after birth without sensory or supraspinal input (Visser, Laurini, de Vries, Bekedam, & Prechtl, 1985). Additional cortico-muscular communication is assumed both with a top-down (Kanazwa, Kawai, Kinai, Iwanaga, Mima & Heike, 2014) as well as a bottom up flow of information (Eyre, 2003). The latter carries sensory information to higher neuronal correlates by two afferent fiber systems. One subserves proprioceptive and tactile information and the second pain, itch, thermocation and specific kinds of touch (Löken et al., 2009). The latter also plays a role in movements directed at one's own body (Bermudez, 2009), which is an essential part of spontaneous movements. Both systems are conceived of together underlying the experience of body awareness (Berlucchi & Aglioti, 2010).

Evidence that neonates specify their body as a distinct entity is assumed from the prominent sensitivity to proprioception in conjunction with other perceptual systems (Bahrick & Lickliter, 2012), temporal synchrony (Fillippetti, Johnson, Lloyd-Fox, Dragovic & Farroni, 2013) and spatial congruence (Rochat & Morgan, 1995). Key elements are the propensity to detect invariant features (e.g. Bahrick & Watson, 1985; Schmuckler, 1996) and proprioception as the "sense modality of the self par excellence" (e.g. Jounen & Gapenne, 1995; Rochat & Striano, 2000). For example, hand-mouth coordination provides neonates with invariant sensory information by bringing regions of their own body in relation to one another (Blass, Fillion, Rochat, Hoffmeyer & Metzger, 1989; Watson, 1984).

Indications of such invariant properties in spontaneous movements come from dynamic approaches studying processes of self-organization. They report fluctuating coordination modes of the limbs (Corbetta & Thelen, 1996) and temporally and spatially synchronized positions of the extremities that act as reference points (Assmann, Thiel, Romano & Niemitz, 2007). The present study investigates the trajectories of the end-effectors hands and feet with regard to self-organizing pattern formation that mediate sensory self-specifying experiences.

The dynamic approach offers suitable concepts and methods to investigate the kinematic data in regard to characteristics of developing body-awareness. It allows from a set of quantitative continuous variables, such as the kinematic data, to detect pattern formation of behavior. Pattern formation arises by self-organization and becomes visible as attractors. A change in the attractor landscape indicates a qualitative change in behavior. New attractors appear due to the dynamic interaction of multiple factors without the controlling function of one superior factor. Characteristics of self-organization are spontaneous attractor formation, phase transitions between different dynamic regimes, and self-similarity on multiple timescales: similar attractor formation on different timescales. The study of self-organization allows to detect features of the intrinsic propensity to create new patterns, which is at the heart of developmental processes.

Based on the characteristics of self-organization, the present study examines the inherent tendency of spontaneous movements to be involved in the development of body awareness: the attraction to certain experiences that provide newborns with invariant sensory information comparable to proprioception in conjunction with other modalities (Bahrick & Lickliter, 2012), temporal synchrony (Fillippetti et al., 2013), spatial congruence (Rochat & Morgan, 1995) and bringing different regions of the body into specific relationships (Blass et al., 1989; Watson, 1984). Two characteristics, which are part in both the concept of self-organization and developing body awareness are also considered: high variability and the organization of behavior as a function of past experience (van der Meer, van der Weel & Lee, 1995; Rochat, 2007; Smith & Thelen, 2003).

Symbolic dynamics is a tool from the dynamic approach that allows to extract robust properties of the dynamics by partitioning of the state space into a finite number of regions (Graben & Kurths, 2003, Hao, 1991). It enables the detection of both the characteristics of self-organization and the interplay of the four effectors with respect to temporal synchrony and spatial congruence and relations.

The present study first displays the investigation of the radial distance trajectories of spontaneous limb movements in regard to criteria of self-organizing pattern formation. Subsequently the results are considered with respect to perceptual characteristics assigned to developing body awareness as well as emotional and cognitive processes.

## 2. Material and methods

*2.1. Data collection*

*2.1.1. Subjects*

Subjects were fourteen healthy full-term neonates from three different periods of data acquisition. The first group consists of seven neonates recruited at a maternity clinic (mean gestagational age: 40.2 weeks, SD: 0.8 weeks; mean age at observation: 5.7 days, SD: 6.8 days; 4 female, 3 male). The second group comprises two neonates recruited before birth by advertisement at the university (mean gestagational age: 40 weeks, SD: 0 weeks; mean age at observation: 45.5 days, SD: 22.1 days; 1 female, 1 male). The third group consists of five neonates recruited at a maternity clinic (mean gestagational age: 39.5 weeks, SD: 1.0 weeks; mean age at observation: 41.4 days, SD: 29.6 days; 4 female, 1 male) (table 1).

Table 1: Overview of infants

Group 1: infants 1-7

| Episode Number | Infant | Age [days] | Sex | Tracking Rate [dp/s] | Data Points | Number of Windows [300dp] Original (surrogate) | Episode Duration [h:mm:ss:dp] |
|---|---|---|---|---|---|---|---|
| 1 | Infant 1 | 1 | female | 5.0 | 3030 | 138 (135) | 0:10:06:00 |
| 2 | Infant 1 | 10 | female | 3.85 | 4654 | 217 (217) | 0:20:08:03 |
| 3 | Infant 2 | 2 | male | 5.0 | 1553 | 64 (63) | 0:05:10:03 |
| 4 | Infant 2 | 4 | male | 5.0 | 1300 | 51 (51) | 0:04:20:00 |
| 5 | Infant 3 | 2 | female | 3.85 | 1664 | 70 (67) | 0:07:12:01 |
| 6 | Infant 3 | 3 | female | 3.85 | 2103 | 92 (89) | 0:09:06:01 |
| 7 | Infant 4 | 1 | female | 3.85 | 2503 | 112 (111) | 0:10:50:01 |
| 8 | Infant 4 | 3 | female | 3.85 | 1670 | 53 (51) | 0:07:13:03 |
| 9 | Infant 5 | 1 | male | 3.85 | 1470 | 53 (51) | 06:21:03 |
| 10 | Infant 5 | 3 | male | 3.85 | 1645 | 60 (59) | 0:07:07:01 |
| 11 | Infant 6 | 1 | female | 4.16 | 1328 | 53 (51) | 0:05:19:01 |
| 12 | Infant 6 | 2 | female | 5.0 | 1272 | 50 (49) | 0:04:14:02 |
| 13 | Infant 7 | 12 | male | 5.0 | 2805 | 127 (121) | 0:09:21:00 |
| 14 | Infant 7 | 26 | male | 5.0 | 1939 | 83 (82) | 0:06:27:04 |
| Mean | 1-7 | 5.7 | | 4.37 | 2067 | 87 (85) | 0:07:53:02 |
| Sum | 1-7 | | 4f; 3m | | 28936 | 1223 (1197) | 1:52:59:02 |

Group 2: infants 8-9

| Episode Number | Infant | Age [days] | Sex | Tracking Rate [dp/s] | Data Points | Number of Windows [750dp] Original (surrogate) | Episode Duration [h:mm:ss:dp] |
|---|---|---|---|---|---|---|---|
| 15 | Infant 8 | 34 | female | 25.0 | 51000 | 1006 (999) | 0:34:00:00 |
| 16 | Infant 8 | 72 | female | 25.0 | 43950 | 865 (861) | 0:29:18:00 |
| 17 | Infant 9 | 22 | male | 25.0 | 42825 | 843 (816) | 0:28:33:00 |
| 18 | Infant 9 | 54 | male | 25.0 | 40225 | 791 (786) | 0:26:49:00 |
| Mean | 8-9 | 45.5 | | 25.0 | 44500 | 876 (866) | 0:29:40:00 |
| Sum | 8-9 | | 1f, 1m | | 178000 | 3505 (3462) | 1:58:40:00 |

Group 3: infants 10-14

| Episode Number | Infant | Age [days] | Sex | Tracking Rate [dp/s] | Data Points | Number of Windows [750dp] Original (surrogate) | Episode Duration [h:mm:ss:dp] |
|---|---|---|---|---|---|---|---|
| 19 | Infant 10 | 26 | male | 50.0 | 29073 | 568 (562) | 0:09:41:23 |
| 20 | Infant 10 | 93 | male | 50.0 | 20968 | 406 (401) | 0:06:59:18 |
| 21 | Infant 11 | 26 | female | 50.0 | 21099 | 408 (401) | 0:07:01:49 |
| 22 | Infant 11 | 74 | female | 50.0 | 37158 | 730 (724) | 0:12:23:08 |
| 23 | Infant 12 | 32 | female | 50.0 | 30928 | 605 (601) | 0:10:18:28 |
| 24 | Infant 13 | 16 | female | 50.0 | 36987 | 726 (724) | 0:12:19:37 |
| 25 | Infant 14 | 23 | female | 50.0 | 29387 | 574 (570) | 0:09:47:37 |
| Mean | 10-14 | 41.4 | | 50.0 | 29371 | 574 (569) | 0:09:47:20 |
| Sum | 10-14 | | 4f, 1m | | 205600 | 4017 (3983) | 1:08:32:00 |
| 1-25 | | | | | | | |
| Mean | 1-14 | 21.7 | | | | | |
| Sum | 1-14 | | 9f, 5m | | | | 5:00:11:00 |

*2.1.2. Measurement Procedure*

For all three groups, the agreement of the ethical commission and parental informed consent was secured in advance. Measurements took place approximately 0,5-1h after and about 1-2h before feeding. Room temperature was set at 30-32°C. During sessions, infants were undressed and allowed to move spontaneously in the supine position. No specific stimuli were presented nor was the spontaneous posture of the infants controlled.

The design of these observations was a frame-by-frame movement microanalysis using videotape recording with synchronized cameras (50Hz) that focused into a volume calibrated by a calibration frame.

Group 1 used three cameras positioned in a distance of 1.5 meters from the baby's feet and 1.5 meters above the ground yielding an observation angle of approximately 45 degrees. Cameras (3) were positioned at the end of the vertical body axis at the feet of the baby and about 45 degrees to the right and left from the first camera circular around the baby. Videotaping was conducted in the maternity clinic in an examination room. If possible, neonates were seen on the first one to four days after birth during their stay in the clinic. Infants 1 and 2 stayed 10 days. Infant 6 visited the lab weekly for 4 weeks after birth due to the mother's interest in the research project. Data were recorded from each infant on two to four separate days, according to their availability. Soft markers with a diameter of 1cm for manual tracking at the forehead, chest, shoulders, elbows, hands, hips, knees and feet were attached prior to videotape recording. For kinematic data analyses, 7 out of 20 infants met the selection criteria: availability of movement episodes from two different days that showed (i) infants in an awake state (ii) no crying and (iii) continuous motor activity longer than 4 minutes without sleeping (eyes closed), crying or resting longer than 20s. Continuous motor activity was defined if one of the limb's end effector exceeded a tangential velocity of 50 mm/s or if the sum of the velocity of the four limbs exceeded 100 mm/s for more than 1s. From the available videotapes of each day and each infant, the longest movement episodes that met the criteria were selected. Due to the time consuming manual tracking of group 1, we did not analyze all available movement episodes: 14 movement episodes between 4 and 20 minutes were analyzed.

Group 2 videotaping was conducted in a movement analysis lab. For a longitudinal study four subjects were recruited at the university and visited the lab at the age of ca. 4, 8, 12 and 16 weeks. Cameras (4) were positioned at the diagonal body axes in a distance of 1.5 meters and a height of 1.5 meters. Passive reflecting markers for automatic tracking in black light were attached at the forehead, chest, navel, hands and feet. Two subjects met the criteria (i)-(iii) described for group 1.

Group 3 data were provided by a lab that used seven synchronized infrared cameras with infrared strobes attached. Passive reflecting markers were affixed to the forehead, chest, upper arms, forearms, hands, tights, calves and feet. Due to storage capacity of storage media, data sampling was interrupted for ca. 15s approximately every 120s for changing the media. Data of 5 infants met the criteria (i)-(iii) and were analyzed to increase the sample and to include some data with a sampling rate of 50 Hz, although the interruption of sampling was not ideal for the sequential analysis with symbolic dynamics.

*2.1.3. Kinematic data*

Kinematic data of 25 movement episodes between 4 and 34 minutes from 14 infants were generated. Movement kinematics for group 1 and 2 infants were generated using the *Ariel Performance Analysis System* (APAS). Xyz-coordinates of the chest (Processus xiphoideus), hands (Caput ulnae) and feet (Malleolus lateralis) were used. Due to the time consuming manual tracking of group 1 infants, tracking rates were reduced: Using the 50 Hz that were provided by the videotapes we skipped either 9, 12 or 13 frames between single data points resulting in sampling frequencies of 5.00, 4.17 and 3.85 Hz. Group 2 infants were tracked automatically with a sampling frequency of 25 Hz and online manual editing, if markers were not seen. Group 3 infants used the optoelectronic 3D motion analysis system *Vicon 370* with automatic tracking at a sampling frequency of 50 Hz. Missing data from periods of unseen markers, were edited manually. Altogether no filters have been used to smooth the resulting trajectories.

Four parameters were calculated from the coordinate data: Radial distance trajectories of the hands and feet from the chest, spatial displacement (difference in effector path distance at time point $t_{i+1}$ from time point $t_i$) and radial displacement (difference in effector distance in relation to the chest at time point $t_{i+1}$ from time point $t_i$) trajectories of the four effectors as well as tangential velocity of the hands and feet (Appendix A.1).

*2.2. Data Analysis*

Data were analyzed in two main steps: First, radial distance trajectories were transformed into a four-dimensional sequence of symbols. In a second step, the distribution of symbols was analyzed on two timescales: first over the entire time series and secondly in shifting windows along the time series.

*2.2.1. Symbolic dynamics*

The first symbolic transformation was applied to radial distance time series of each effector, by using a static two symbol coding ('0' = proximal and '1' = distal) to capture the positions of each of the four effectors $e_j$ (j=1,2,3,4) at each sample point $t_i$ (i=1,2,…,n). To each value of the radial distance time series $R_{ij}$ of one effector, a symbol $s_i$ was assigned in the following way:

(5) $$s_i = \{0: R_j(t_i) < \theta \, ; \, 1: R_j(t_i) \geq \theta\},$$

where the threshold $\theta$ was defined by the radial distance from the chest to some percentage of limb length for each limb respectively: for hand movements, we defined the radial distance from the chest to the point of 50% of arm length as threshold $\theta$ and for feet movements, we used 70% of leg length as threshold $\theta$. These values were assigned as limit between two states (figure 1).

Figure 1

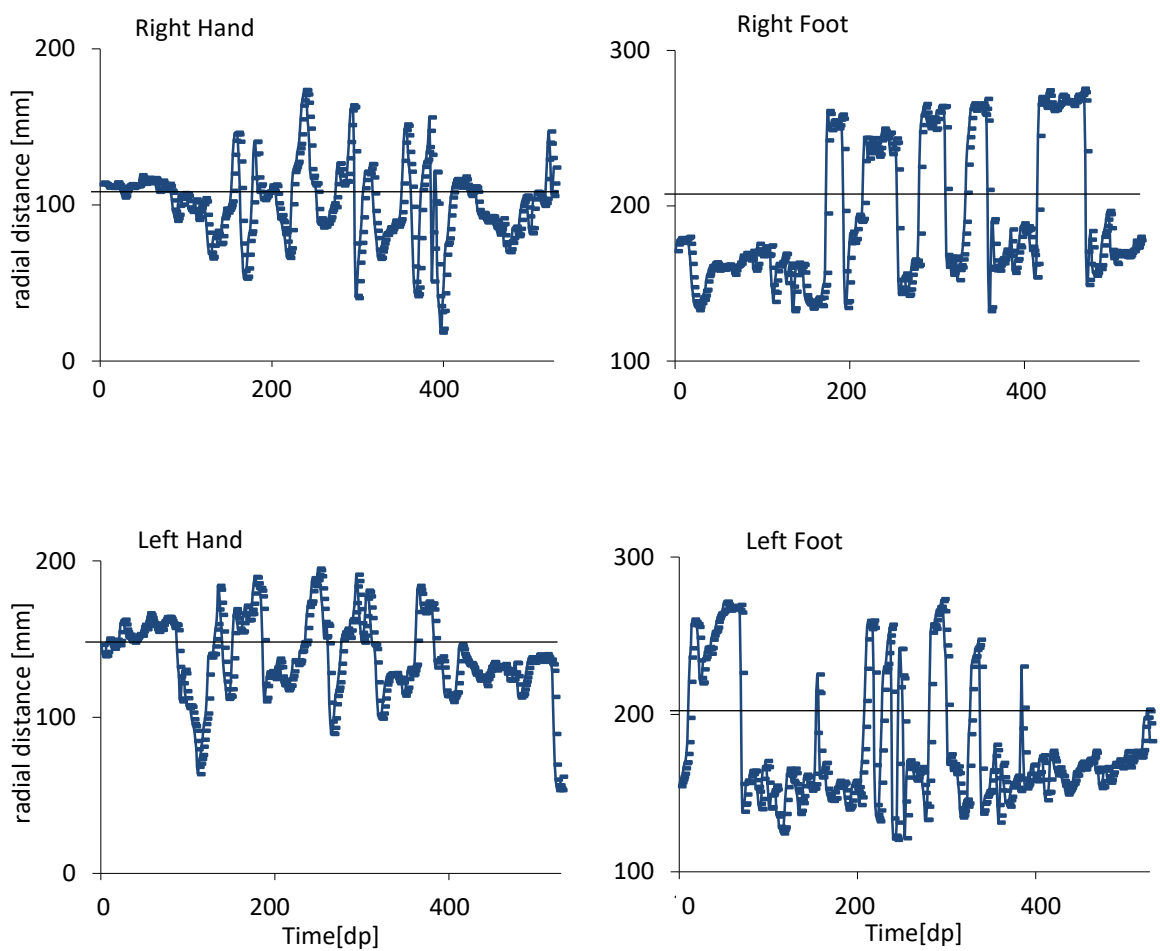

*Figure 1:* Illustration of the first step of the symbolic transformation. Data points 1-500 of the radial distance time series of the end effectors hands and feet of episode 3 (infant2/day2). Horizontal lines denote the thresholds θ for defining the symbols: values below the limits are transformed into the symbol '0'; data points above the horizontal lines obtain the symbol '1'.

'0' refers to a position proximal to the defined body center, '1' refers to a distant position of the end effector. From this procedure resulted four symbolic strings, one for each effector, consisting of a sequence of the single effector symbols 1 and 0. The threshold θ was determined for each effector and each individual movement episode separately.

The parallel processing of the four strings with two alternative states respectively yielded a four dimensional symbolic sequence with each data point consisting of four single effector symbols: e.g. 1010 for a distal state of the right hand (1), a proximal state of the right foot (0), a distant position of the left hand (1) and a close position of the left foot (0). The new four-dimensional time series consisted of $4^2 = 16$ different states, termed body configurations (in the order: right hand, right foot,

left hand, left foot: 0000 = 1, 0001=2, 0010=3, 0100=4, 1000=5, 0011=6, 1100=7, 0101=8, 1010=9, 1001=10, 0110=11, 0111=12, 1011=13, 1101=14, 1110=15, 1111=16) (figure 2).

Figure 2

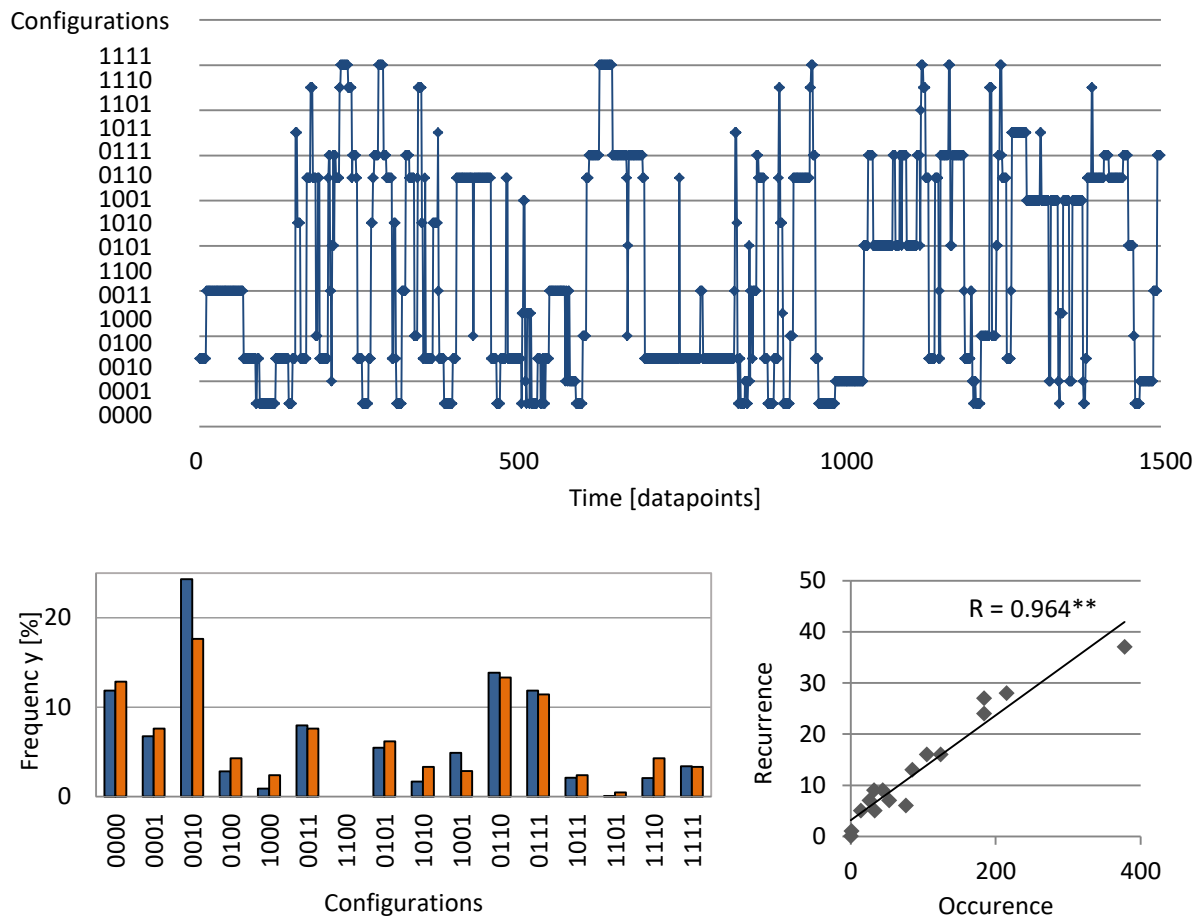

*Figure 2:* Illustration of the second step of the symbolic transformation. Top: Sequence of configurations derived via symbolic transformation of episode 3 (infant2/day2, n=1553, 5Hz, see figure1,3). Bottom: left: Corresponding histogram of occurrence (blue) and recurrence (orange) of the 16 configurations. Right: Correlation diagram derived from plotting occurrence against recurrence. ** the correlation (Pearson) is significant at a level of $\alpha = 0.01$.

For further investigation, the 16 configurations were assigned to 5 classes according to the number of proximal and distal positions: (I) all effectors proximal (1), (II) three effectors proximal, one effector distal, (2-5), (III) two effectors proximal, two effectors distal (6-11), (IV) one effector proximal, three effectors distal, (12-15), (V) all effectors distal (16).

*2.2.2. Distribution of configurations*

The emergence of attractor states was investigated by estimating occurrence and recurrence for each configuration. Occurrence is defined by the cumulative time of staying in a configuration, which is the sum of all data points that display this configuration. Recurrence is defined by the number of times a pattern occurred, regardless of the duration of each occurrence. Significant high values of both parameters for a certain configuration in a predefined time window denote an attractor state.

For the investigation of attractor formation on different timescales, these two measures were computed on two time scales: First for each of the 25 movement episodes for the entire episodes, and secondly for each of the 25 movement episodes in sliding windows along the symbolic time series with a window size of 300/750dp and a shift size of 20/50dp according to the sampling frequency (table 1). A window size of 300dp for group 1 infants corresponds to 60s for a tracking rate of 5Hz and 78s for a tracking rate of 3.85Hz. A window size of 750dp corresponds to 30s for group 2 infants (tracking rate 25 Hz) and 15s for group 3 infants (tracking rate 50Hz).

For statistical analysis, binomial *z*-scores were calculated for the distribution of each of the 16 configurations $x_k$, ($k=1,2,…,16$) versus uniform distribution $x_u$ according to the formula:

(6) $$z = p_1 \text{-} p_2 / [p(1-p)(1/n_1+1/n_2)]^{1/2},$$

with $p_1=x_k/n_1$ and $p_2=x_u/n_2$ being the relative frequencies of the absolute distributions $x_k$ and $x_u$ and $p=(x_k+x_u)/(n_1+n_2)$. For calculation of occurrence: $n_1=n_2=$ number of data points of the episode; for calculation of recurrence: $n_1=n_2=$ number of recurrences of all 16 configurations. With a significance level of $\alpha \leq 0.05$, the critical *z*-score was 1.96. Effect sizes were calculated and according to Hattie (2009), values of Cohen's $d \geq 0.4$ were interpreted as desired effects.

Binomial scores were also used to test the statistical significance of deviations of the frequency distribution of a configuration $x_{ik}$ in a sliding window *i*, ($i=1,2,…,n$) from the distribution of the overall movement episode $x_k$. Binomial *z*-scores were computed according to formula (6), with $p_1=x_{ik}/n_1$ and $p_2=x_j/n_2$ being the relative frequencies of the absolute distributions $x_{ik}$ and $x_k$, $p=(x_{ik}+x_k)/(n_1+n_2)$. For calculation of occurrence: $n_1=$ number of data points of a window and $n_2=$ number of data points of the episode; for calculation of recurrence: $n_1=$ number of recurrences of all 16 configurations in a window and $n_2=$ number of recurrences of all 16 configurations in the episode. The resulting *z*-scores of the original time series were compared with the 0.05 and 0.95 quantiles of *z*-scores obtained from surrogate data by the same procedure.

*2.2.3. Surrogate Data Analysis*

Surrogate time series were created by the Iterated Amplitude Adjusted Fourier Transform (IAAFT) from the original time series by the program "surrogates" from the software package TISEAN (Hegger, Kantz & Schreiber, 1999; Schreiber & Schmitz, 2000). Surrogates correspond to the null hypothesis of a linearly correlated stochastic Gaussian process. Due to the fourier transformation, surrogate time series are occasionally shorter than the original time series (table 1).

# 3. Results

## 3.1. Relation between spatial and radial displacement

To proof the significance of the parameter radial distance for the symbolic transformations and subsequent analyses, the correlation between spatial and radial displacement of each effector was calculated (figure 3). The mean over all 25 episodes for the right hand is R = 0.76, for the right foot R = 0.87, for the left hand R = 0.79 and for the left foot R = 0.87, with significant correlations (Pearson) for all episodes at the $\alpha = 0.01$ level (data not shown).

Figure 3

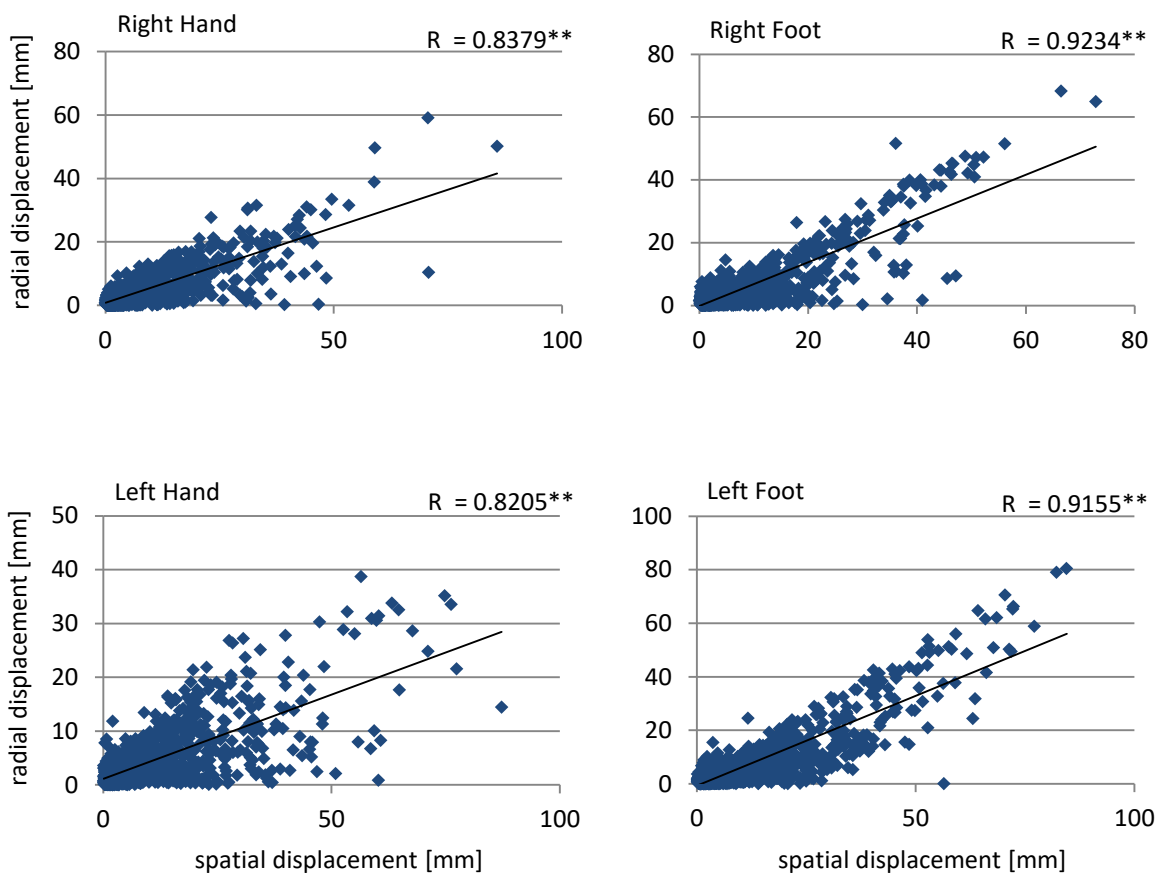

*Figure 3:* Illustration of the correlation between spatial and radial displacement of episode 3 (infant2/day2, n=1553, 5Hz). **correlations (Pearson) are significant at a level of $\alpha = 0.01$.

The high correlations show that both measures use the same point as reference. This means that limb movements are directed in reference to the body center and that the parameter radial distance captures the crucial dynamics of the limb movements.

*3.2. Attractor formation: Evaluation of the distributions of configurations in movement episodes*

For the investigation of attractor-states the distributions of the configurations were analyzed by means of the parameters occurrence and recurrence. First, occurrence and recurrence were highly correlated as illustrated in figure 2. This accounts for all episodes: All correlations are statistically significant at α = 0.01 level (data not shown). The average correlation pooled over all movement episodes is R = 0.92. This shows that those configurations, in which infants were staying for the longest cumulative time, were the same configurations, to which they returned with the highest frequency.

Second, episodes were investigated for attractor states. Attractor states were defined by significant higher values of occurrence and recurrence according to binomial test against uniform distribution ($\alpha \leq 0.05$; $d \geq 0.4$). 12 out of 14 infants show 1-3 configurations that act as attractors in one or both episodes (mean: 1.59, SD: 0.77) (figure 4).

Figure 4

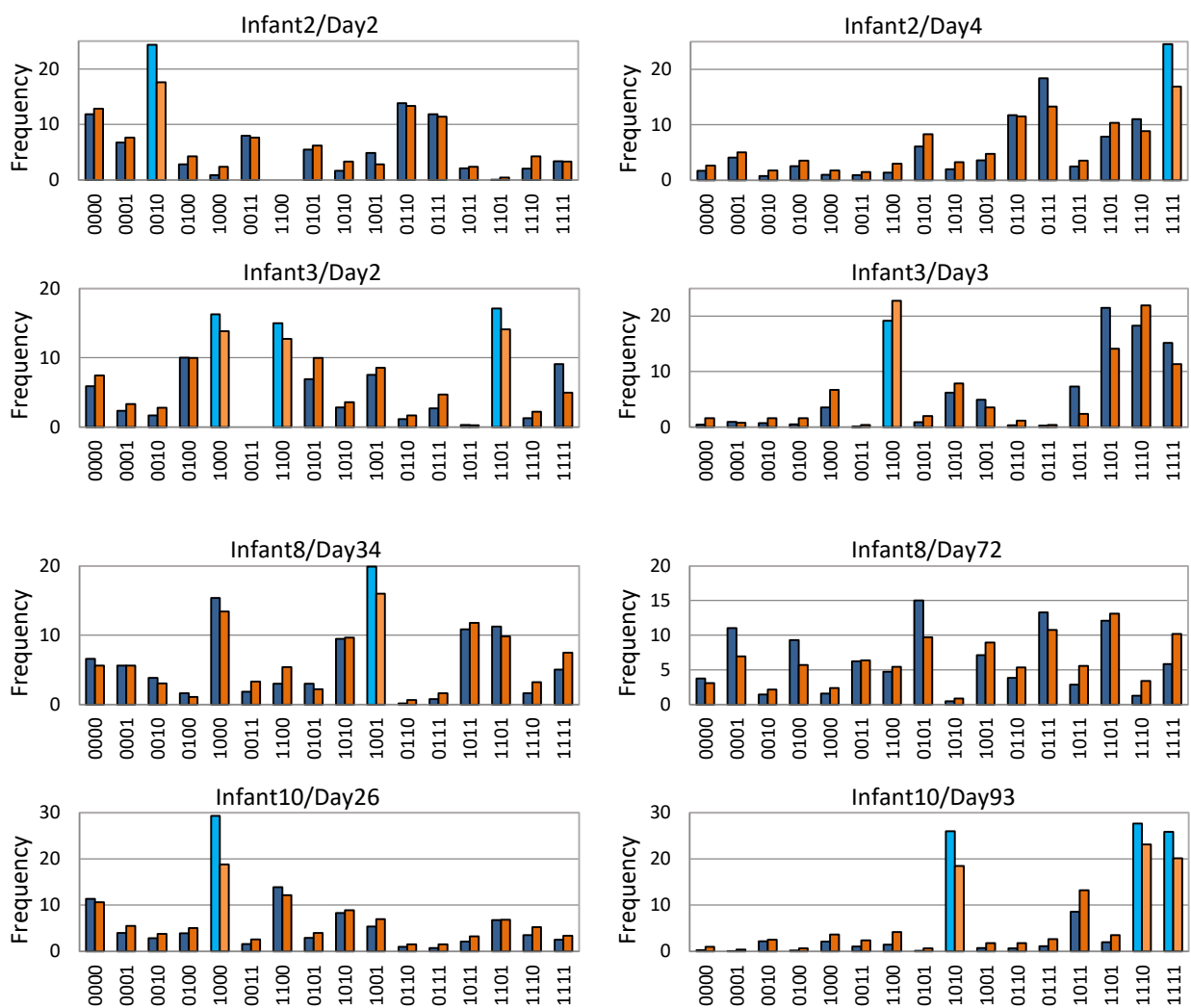

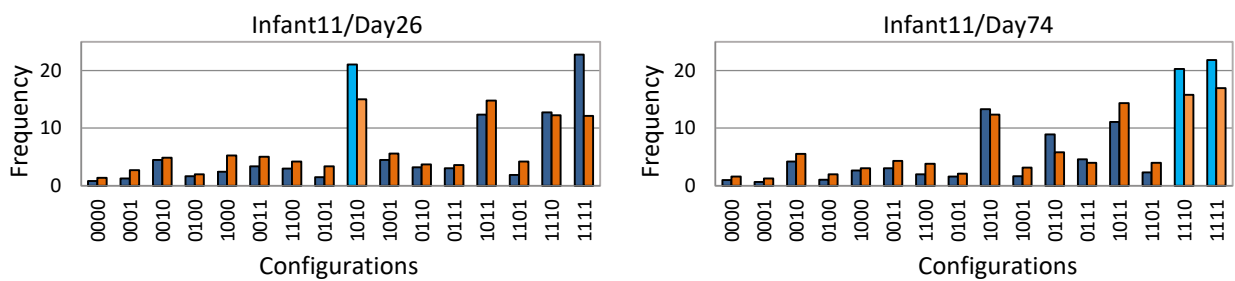

*Figure 4a*: Relative frequency distributions measured in percent of occurrence (blue) and recurrence (orange) of selected movement episodes. Horizontal histograms correspond to episodes of the same infant on two different days respectively. Light fillings mark significant high values of both parameters estimated by binomial testing against uniform distribution ($\alpha = 0.05; d \geq 0.4$).

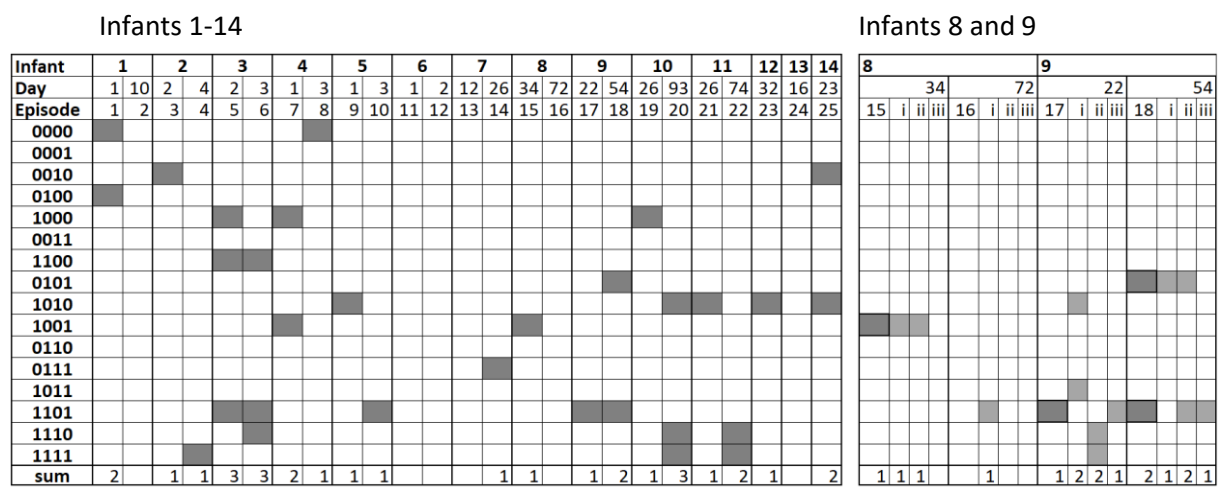

*Figure 4b*: Pattern of preferred configurations for all 14 infants/25 episodes. Filled segments represent favored configurations that act as attractors.

Seven infants have attractor configurations in two episodes, five in one episode and two in none. This indicates, that on this timescale of episodes (4-34 min) infants switched between present and absent preferred configurations as well as the quality of the favored configurations from one episode to the next. This is also confirmed by very low correlations (except the outlier of infant 11) of the distributions of occurrence and recurrence between two episodes of the same child (mean (occurrence): 0.153, mean (recurrence): 0.198; table 2a).

Table 2a: Correlations of occurrence and recurrence between subsequent movement episodes of individual infants

| Episodes | Infant | Occurrence | Recurrence |
| --- | --- | --- | --- |
| 1/2 | Infant 1 | 0.295* | 0.357* |
| 3/4 | Infant 2 | 0.000 | 0.000 |
| 5/6 | Infant 3 | 0.228 | 0.110 |
| 7/8 | Infant 4 | 0.093 | 0.343* |
| 9/10 | Infant 5 | 0.011 | 0.043 |
| 11/12 | Infant 6 | 0.078 | 0.212 |
| 13/14 | Infant 7 | 0.026 | 0.016 |
| 15/16 | Infant 8 | 0.028 | 0.000 |
| 17/18 | Infant 9 | 0.124 | 0.230 |
| 19/20 | Infant 10 | 0.011 | 0.004 |
| 21/22 | Infant 11 | 0.787** | 0.859** |
| Mean | Infants 1-11 | 0.153 | 0.198 |
| SD | Infants 1-11 | 0.231 | 0.258 |

\* significant at a level of $\alpha = 0.05$
\*\* significant at a level of $\alpha = 0.01$

Preferred configurations correspond to attractor-states in self-organizing systems that attract the behavior of the system. Switches show, that attractors emerge spontaneously, dissolve, and new attractors emerge. This corresponds to phase transitions between different dynamical regimes: periods with attractor states, periods without attractor states as well as changes between different attractor states.

The long episodes of infants 8 and 9 (mean: 29:40:00 min) were additionally analyzed in three segments of about 10 minutes (i, ii, iii; figure 4b) and show that emergence, dissolving and transition of attractor states emerge on both time scales (10min and 30min): this indicates self-similarity as a feature of self-organization. The patterns of the overall episodes are reflected quite well in the 10 minutes sections, with 20 minutes-time periods showing identical present or absent preferred configurations as the overall episode. This is also reflected by correlations of occurrence and recurrence between the sections i/ii, ii/iii and i/iii (table 2b). This implies that organizational units may span time frames of 20 minutes and that the shorter episodes of 5-10 minutes in this study may be representative for adjacent time periods with altogether up to 20 minutes.

Table 2b: Correlations of occurrence and recurrence between subsequent sections of the movement episodes of infants 8 and 9

| Infant/day/episode | Section | Occurrence | Recurrence | MW |
|---|---|---|---|---|
| Infant 8/ day 34/ episode 15 | i/ii | 0.806** | 0.936** | |
| | ii/iii | 0.375* | 0.664** | |
| | i/iii | 0.449** | 0.712** | 0.657 |
| Infant 8/ day 72/ episode 16 | i/ii | 0.649** | 0.590** | |
| | ii/iii | 0.101 | 0.381* | |
| | i/iii | 0.181 | 0.560** | 0.410 |
| Infant 9/ day 22/ episode 17 | i/ii | 0.083 | 0.410** | |
| | ii/iii | 0.351* | 0.729** | |
| | i/iii | 0.010 | 0.173 | 0.293 |
| Infant 9/ day 54/ episode 18 | i/ii | 0.627** | 0.590** | |
| | ii/iii | 0.921** | 0.862** | |
| | i/iii | 0.726** | 0.287* | 0.669 |
| Mean | | 0.440 | 0.575 | |
| SD | | 0.306 | 0.228 | |

\* significant at a level of $\alpha = 0.05$
\*\* significant at a level of $\alpha = 0.01$

To investigate, if certain configurations are preferred because of primitive reflex activity or biomechanics due to homology or bilateral symmetry, figure 5 displays the distribution of all 16 configurations both pooled over all infants as well as over each age category. There are no significant *z*-scores of any configuration reaching an effect size of $d \geq 0.4$ in both occurrence and recurrence. This shows, that infants used all configurations with a comparable magnitude without preference for any configuration and irrespective of their temporary attractor-function. For example, there are no preferences of the configurations 6 (1100) and 7 (0011) corresponding to the asymmetric tonic neck reflex (ATNR), or 8 (0101) and 9 (1010) according to homology. That means that primitive reflex activity or biomechanics due to homology or bilateral symmetry are no prevailing determining factors and points to alternative organizational mechanisms or factors, e.g. perceptual processes, from which attractors emerge.
15

Figure 5

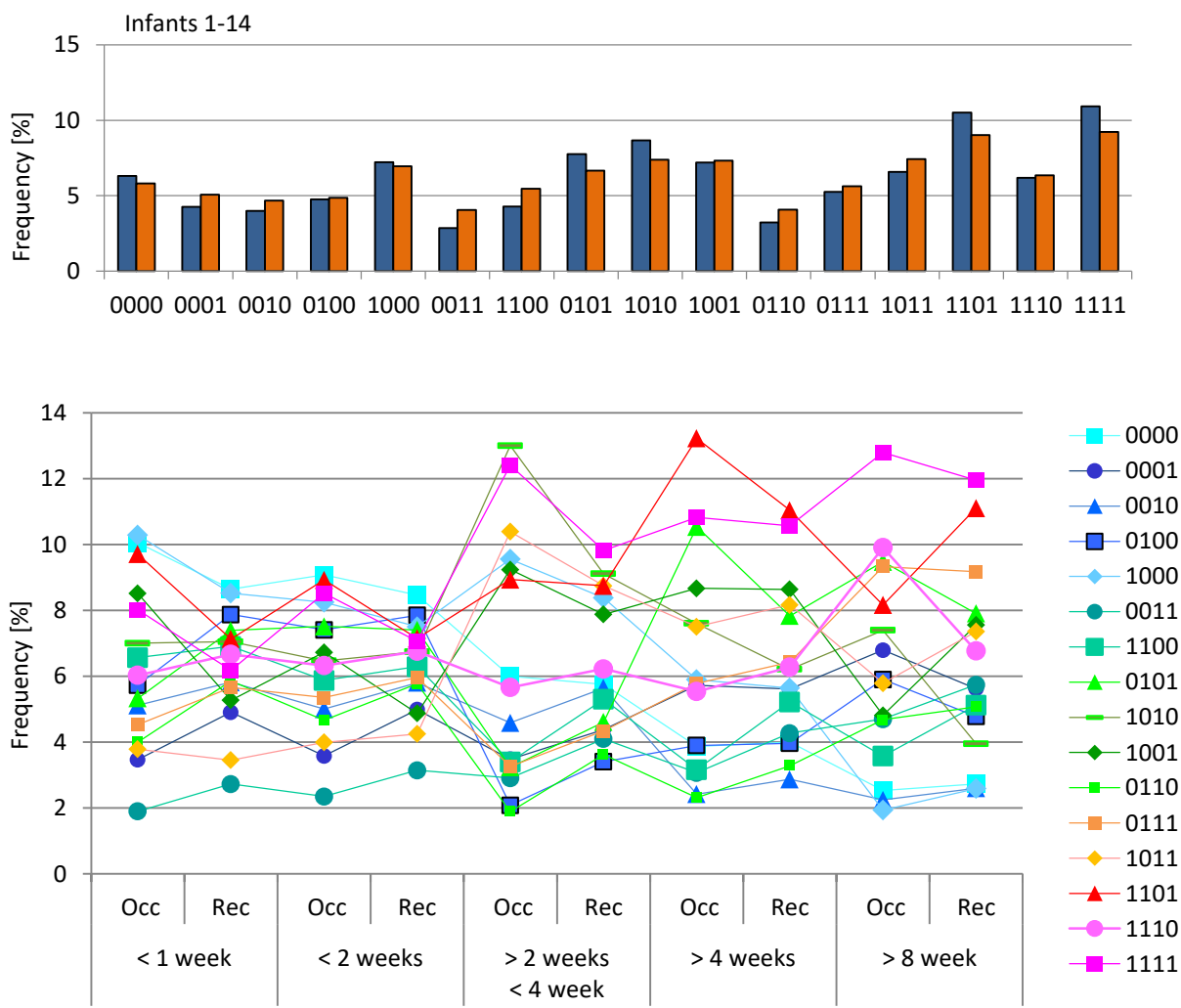

*Figure 5*: Top: Relative frequency distributions of occurrence (blue) and recurrence (orange) pooled over all 14 infants/25 episodes. Bottom: Relative frequency distributions of occurrence and recurrence for infants of different age categories (see A.2., databases of figures).

With increasing age, figure 5 below shows an increase of configurations with more distal positions of the effectors and a decrease of configurations with more proximal positions to the body center with a shift around two weeks. There is no hint towards a prevailing significance of the dimensions along the body axis, cranial-caudal and right-left.

*3.3. Evaluation of the distributions of classes of configurations in movement episodes*

For a closer examination of the central-peripheral dimension, the configurations were divided into five classes according to the number of proximal and distal effector positions (figure 6). There is no significant preference for one class for all children. The distributions for the different age groups confirm the increase of distal and decrease of proximal configurations with age as well as the shift at the age of two weeks. Testing against uniform distribution revealed that infants > 4 and 8 weeks significantly preferred class V (configuration 1111).

Testing individual age groups against each other shows significant differences between children < 1 and 2 weeks and children > 4 and 8 weeks for class I and between children < 1 week and > 8 weeks additionally for class V. This shows quantitatively the transition from writhing to fidgety movements that has been described qualitatively between 6-9 weeks (Einspieler, Bos, Libertus & Marschik, 2016). These results indicate that the change may begin around 2-4 weeks, and that the proximal-distal dimension plays a significant role. There is a change in orientation from inwards to outwards in direction to the peripersonal space.

Figure 6

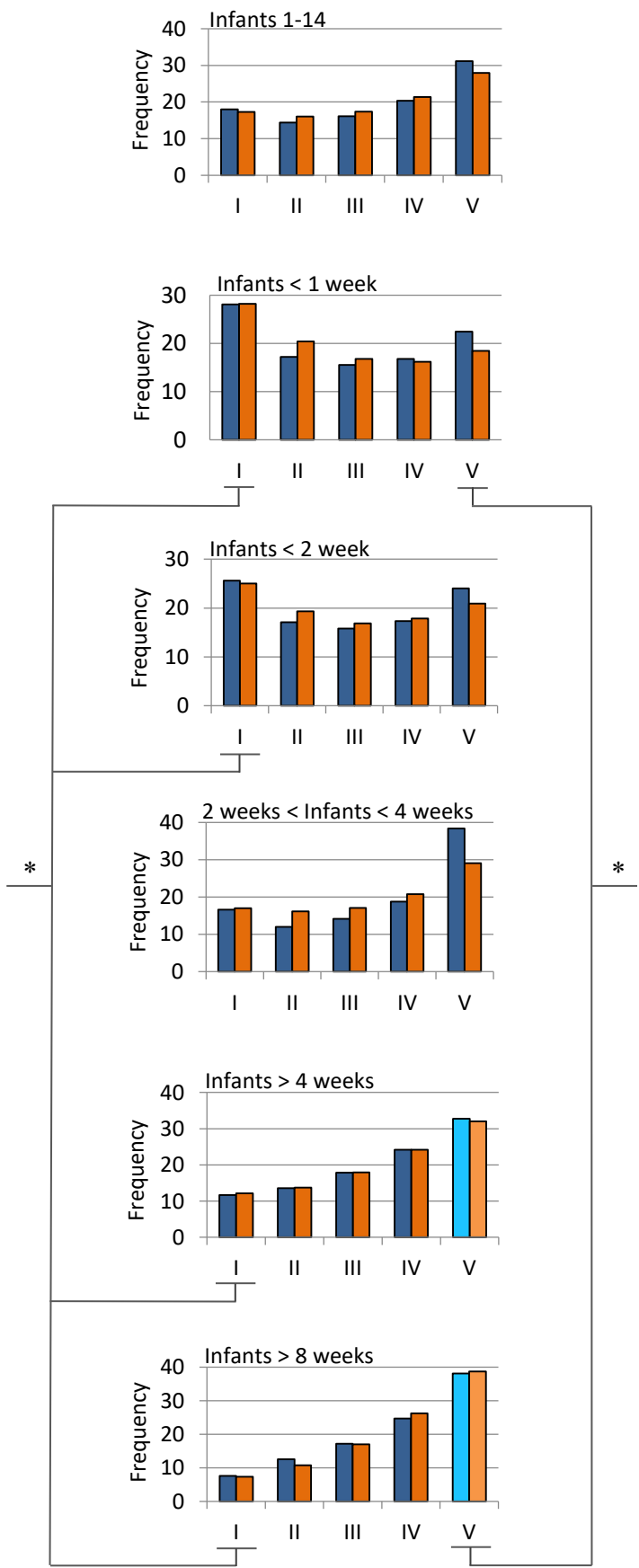

*Figure 6*: Relative frequency distributions of occurrence (blue) and recurrence (orange) specified in per cent of the 16 configurations combined into 5 classes. Values of occurrence and recurrence are normalised according to the number of configurations in each class. Light fillings mark significant high values within distribution; * mark significant differences between distributions ($\alpha \leq 0.05$; $d \geq 0.4$). (see A.2., databases of figures).

*3.4. Testing for self-similarity: Emergence of attractors on a lower timescale*

The distributions of occurrence and recurrence were recorded on a lower time scale in sliding time windows along the movement episodes. Significance was proved by calculating binomial scores against the corresponding distribution throughout the episode (Figure 7). *Z*-scores of individual configurations above the 0.95 quantiles of the surrogates indicate nonlinearities that show sequential dependencies that go beyond the linear correlations of the surrogates and emerge as preferred configurations. Both parameters above the significance level indicate attractor states. The superimposed representation in the lower part of figure 7 shows self-similarity with the pattern of preferred configurations on the higher time scale of episodes.

Transitions between qualitatively distinct dynamical regimes are criteria of self-organization and are shown by periods with linear correlations without preferred configurations and non-linear periods with preferred configurations. This self-similar pattern formation on different time scales demonstrates how the amplification of fluctuations on lower levels lead their way to observable "qualitative change" on higher levels.

Figure 7

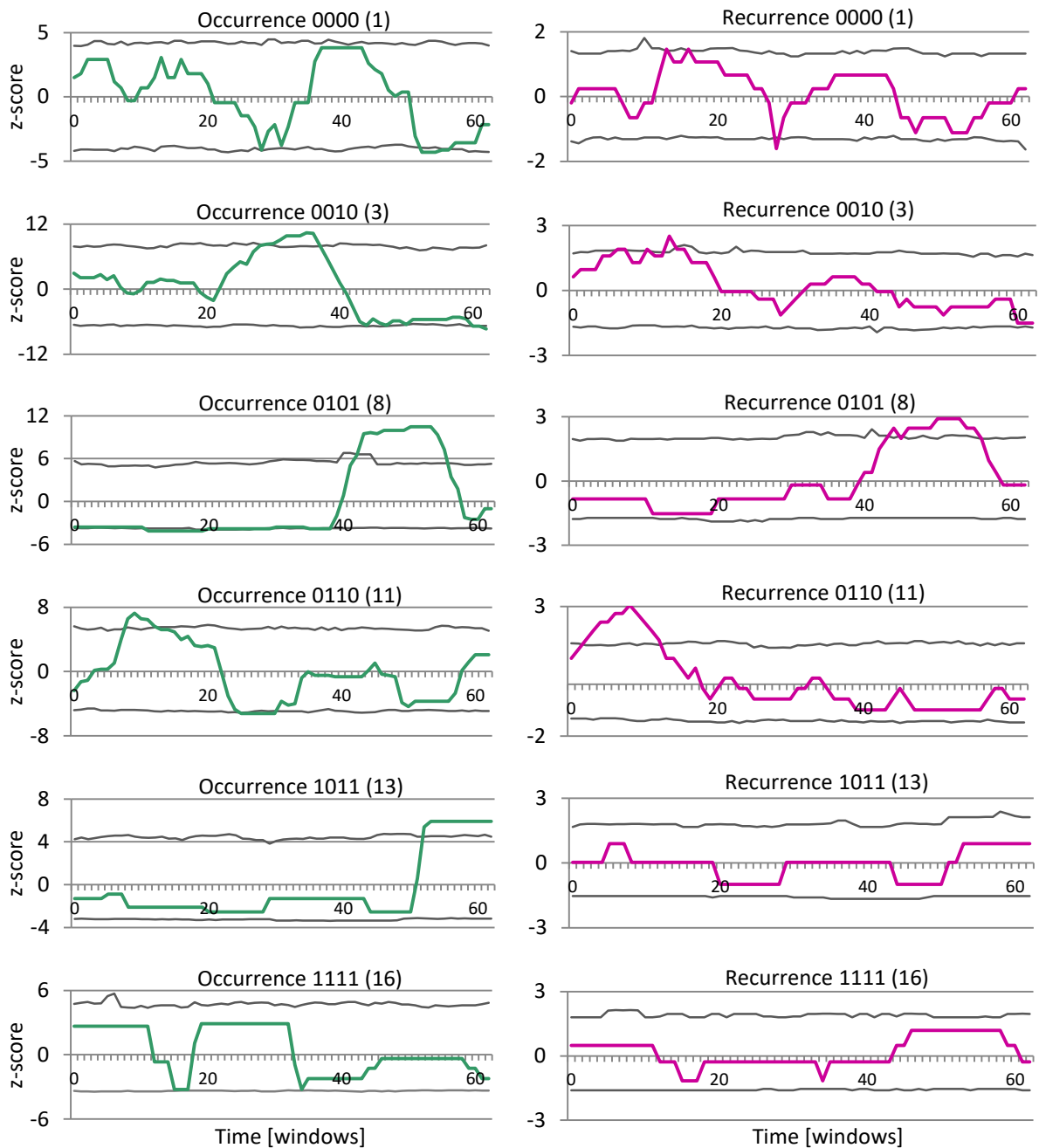

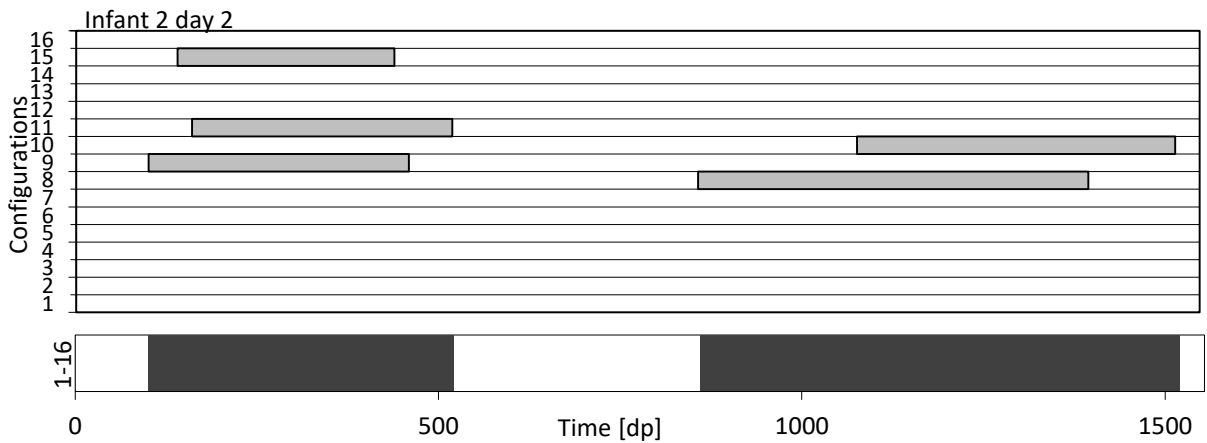

*Figure 7*: Illustration of binomial scores of occurrence and recurrence of exemplary configurations in sliding windows against the distribution of the entire episode 3 (infant2/day2, see fig.1,2,3; window size: 300dp=60s/ step size: 20dp=4s). Green and pink lines present *z*-scores of the progression of the original time series of occurrence and recurrence. Grey lines display 0.05 and 0.95 quantiles of *z*-scores obtained by the same procedure with 1000 surrogate time series. Bottom: Diagram of attractor configurations. Light grey: time sections with significant higher values of both parameters. Dark grey: resulting pattern of alternating sections with (grey) and without (white) significant configurations by super positioning of the upper light bars.

Figure 8 displays the attractor formation on the lower time scale for selected episodes. They confirm the self-similar pattern formation with an alternation between segments with different combinations of preferred configurations and interval segments.

The robustness of the pattern formation, is confirmed by three considerations: First, gaps in the episodes due to sampling procedure for infants 10 and 11 (see methods) occur both within dark and white sections as well as between them, confirming the robustness of the patterns rather than pointing to different patterns before and after the gaps. Second, different tracking rates and used time windows (see methods) show equal patterns in figure 8. Third, increasing window- and shift sizes (window/shift) of group 2 infants (data rate 25 Hz) from 750/50dp (=30/2s) to 1500/100dp (=60/4s) and of group 3 infants (data rate 50 Hz) from 750/50dp (=15/1s) to 1500/100dp (=30/2s) and 3000/200dp (=60/4s) resulted in consistent significant configurations (data not shown).

Figure 8

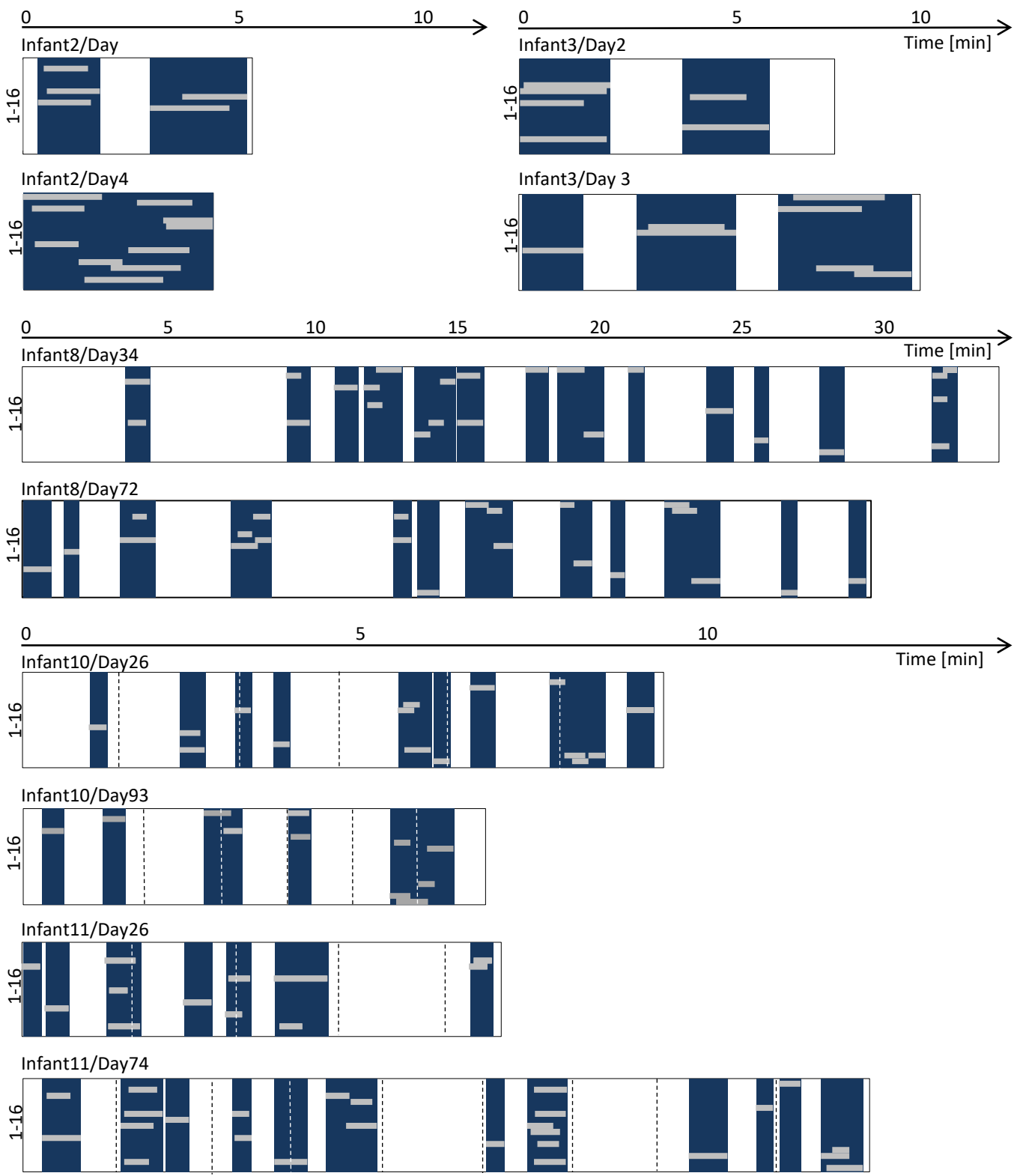

*Figure 8:* Schematic representation of lower time scale patterning within movement episodes (see fig.4a). Dark sections indicate the presence, white intervals the absence of attractor configurations. Light grey bars indicate the composition of the significant individual configurations 1-16. Timelines

refer to mappings below (window/shift size: infant 2: 60s/4s; infant 3: 78s/5.2s; infant 8: 30s/2s; infants 10/11: 15s/1s).

*3.5. Analysis of attractor states*

Occurrence and Recurrence of attractor function was determined for each configuration to investigate to which extent configurations act attractors. Figure 9 shows that all configurations occur as attractors on one or both of the analyzed timescales. This first confirms the results of figure 5, secondly points to a dynamic multicomponent organizational mechanism that goes beyond reflex activity and biomechanical features, and third implicates that the effector endpoints reflect an overall state of body dynamics rather than acting singly or in pairs. In children < 2 weeks a tendency towards configurations with proximal effectors is observable and in children > 2 weeks a reverse tendency. This is in agreement with the results of Figures 5 and 6, and indicates a significance of the proximal-distal dimension in the motor development at two weeks.

Figure 9

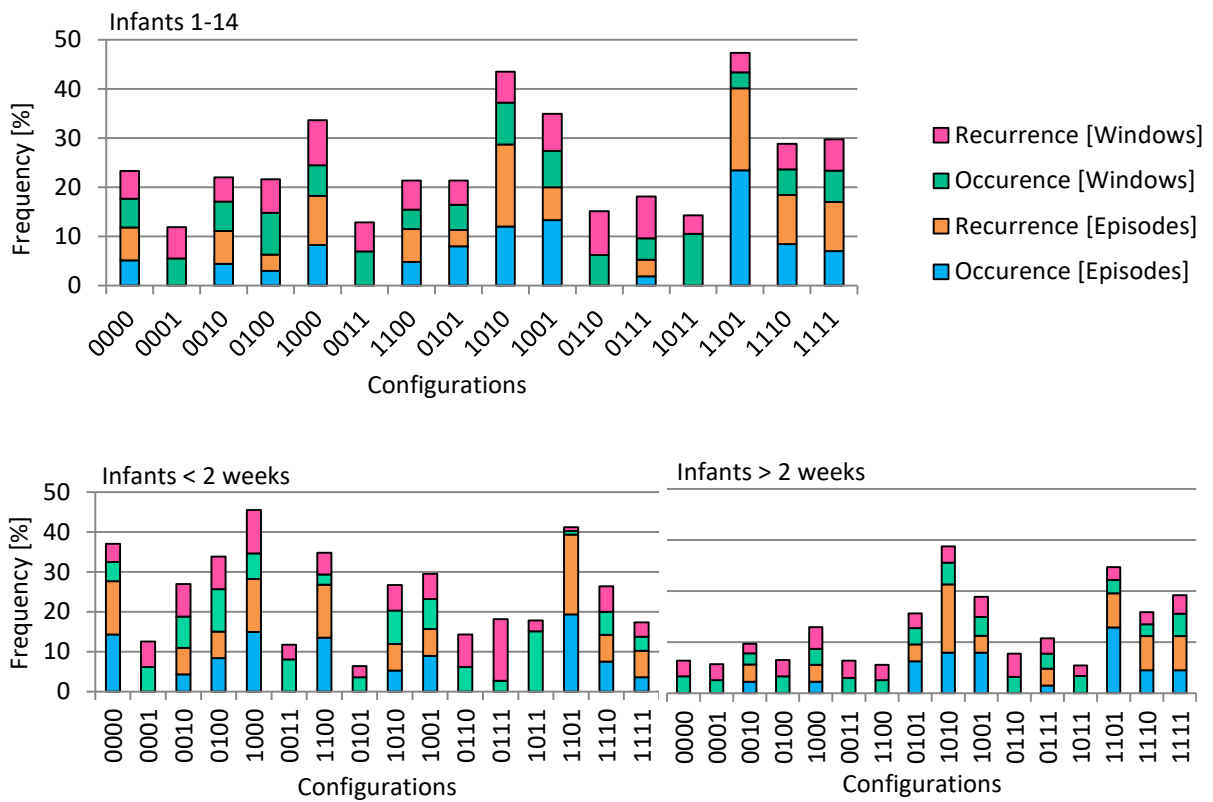

*Figure 9*: Occurrence and Recurrence of the emergence of attractor states of each configuration on the timescales of episodes (4-34min, see fig.4b) and windows (15-60s in steps of 2-5s, see fig.7,8). (see A.2., databases of figures).

## 4. Discussion

The results of the investigation of spontaneous movements with symbolic dynamics show the hypothesized features of self-organization: phase transitions between different dynamical regimes, attractor formation and self-similarity were manifested by the transient emergence of preferred configurations on two time scales. In certain time windows, significantly high values of occurrence and recurrence showed that certain configurations acted as starting and target configurations of movement sequences. Periods without attractor formation as well as the function of all 16 configurations as attractors correspond to the high variability required for self-organizing processes. This alternation between different dynamical regimes on multiple time scales is consistent with results from recurrence analysis (Assmann et al., 2007). Concerning underlying mechanisms of the transient emergence and disappearance of attractor-configurations, dynamic interactions of spontaneous neurogenic and reflexive activity in interaction with the biodynamic properties of the body in gravity have been discussed within the concept of chaotic itinerancy (Assmann, 2008; Tsuda, 1991). The present study shows that all configurations both occur and act as attractors in comparable magnitudes. This follows the dynamic assumptions and points to other influencing factors than primitive reflex activity and biomechanical biases due to homology or bilateral symmetry. The following section examines to what extent perceptual and emotional processes in the development of body awareness might be involved in attractor formation.

*4.1. Perceptual Processes*

*4.1.2. Amodal information and perceptual narrowing*

The attraction to a configuration can be understood as a process of perceptual narrowing (Lewkowicz & Ghazanfar, 2009). Infants are born into a multisensory world with a developing, highly plastic nervous system. With a special sensitivity to invariant amodal attributes, they can discover multisensory stimulation that is both unitary and relevant to them and ignore less relevant stimulation (Gibson, 1969; Lewkowicz, 2002; Lickliter & Bahrick, 2000). Amodal information is invariant across the senses and constitutes a higher-order, global feature, in that it requires the abstraction of information across different modalities over time. With infants' ability to recognize these higher-order relationships, they are able to integrate meaningful multisensory information within a constant flow of sensory signals. Amodal information attracts the attention of the perceiver and is comprised of changes in intensity or temporal and spatial aspects of stimulation such as temporal synchrony, rhythm and tempo (Bahrick, Lickliter & Flom, 2004; Bahrick & Lickliter 2002, 2012). We assume that similar mechanisms to the sensitivity to amodal information are involved in the transient patterns of attractor configurations.

*4.1.1. Temporal Synchrony*

The most fundamental type of amodal information is temporal synchrony. A crucial feature of recurring configurations is temporal synchrony of single limb movements into a limited number of spatial positions, i.e. the proximal or distal position of the end effectors. Due to the biomechanical spring-like properties of the limbs, movement trajectories of the end effectors reverse at certain points.

Movements at the reversal points are characterized by a velocity minimum and therefore more time is spent in these positions. Consequently, synchronization of the turning points of the four limbs is favored, results in the observed reference configurations and mediates synchronous stimuli. Synchrony falls out of the system's dynamics as a "byproduct". Recurrence analysis also demonstrated that movement episodes and pauses of the four limbs are synchronized (Assmann et al., 2007). Onset and offset of movements provide synchronous temporal proprioceptive, kinesthetic and tactile stimulation from two or more moving limbs.

*4.1.2. The intersensory redundancy hypothesis*

The intersensory redundancy hypothesis highlights the importance of temporal synchrony of proprioceptive in conjunction with other sensory stimulation for early perceptual development (Bahrick & Lickliter, 2002, 2012). It has been shown, that infants can detect intersensory synchrony when related to their own body from the first days of life. This early sensitivity to the timing of events rests on relative timely dependencies of two events up to temporal delays of 5 seconds than on precise temporal coincidences (Fillippetti et al., 2013). This suggests that the recurrence to four effector configurations, in which limbs move in relative rather than precise synchrony, falls within the window of infants' sensitivity to temporal synchrony and are recognized as salient stimuli.

Perceptual sensitivity to a stimulus draws attention to this stimulus and induces orienting behavior. The salience of synchronous events and amodal stimulation plays a crucial role in attention allocation. Infants are effective in establishing patterns for selective attention to relevant and coherent aspects of their experiences. Selective attention to self-generated stimulation by exploratory behavior is considered as playing the key role of what is perceived, learned and memorized. In turn, this influences the focus of attention in following bouts of exploration. Since exploratory activity is the source for new stimulation, action is tightly coupled with these ongoing feedback-loops (Adolph & Berger, 2005; Gibson, 1988; von Hofsten, 1993). This implies that synchronizations of certain inflection points of the four effectors produce synchronization experiences that draw selective attention to the corresponding configuration and favors the recurrence to this state. This way, the above-described cycles of action, attention, perception, learning, and memory, lead to the observed recurring patterns of particular configurations.

From a dynamic perspective the feedback loops of action, attention, perception, learning and memory can be conceived of as a system of interacting dynamic influences that evolve over time on a behavioral and experiential level as well as within the organization of neuronal correlates. Dynamic developmental theories highlight the importance of regressive processes (Adolph & Berger, 2005; Gibson, 1988; Lewis, 2000; Thelen & Smith, 1994). The various subparts of a system begin to cooperate with each other and assemble into more stable and efficient patterns. The recurrence to a certain configuration is a process of reducing degrees of freedom (Bernstein, 1967) and perceptual narrowing (Lewkowicz & Ghazanfar, 2009). The various subparts are the four effectors that begin to cooperate with each other and assemble into preferred recurrent configurations that provide stability.

Dynamic approaches also emphasize the influence of developmental history and experience that narrows the perceptual sensitivity (Smith & Thelen, 2003). The literature on the development of body awareness refers to the organization of behavior as a function of past experience (Rochat, 2007). This aspect can be found in the observed recurrence of certain configurations. What is experienced and familiar is attractive and reproduced. This is also in line with the idea of selective attention and the feedback loops of action, attention, perception, learning, and memory.

In recurrent configurations, the limbs are brought into certain recurrent spatial relations to each other. Even before birth, infants are observed to bring body parts in direct relation to one another in hand-mouth coordination (Blass et al., 1989; Watson, 1984). This provides neonates with invariant sensory information that specifies their own body as a bounded entity (Rochat 2012). Likewise, the timing of limb movements into certain recurrent configurations provides infants with similar self-specifying invariant sensory experiences. This way they are involved in the differentiation of the self from others and the formation of a coherent body awareness. Infants appear sensitive to self-specifying invariants as provided by proprioception in conjunction with other modalities from birth onwards (Rochat & Striano, 2000).

*4.1.3. Synchronous proprioceptive stimulation*

In addition to synchrony across different sensory modalities we postulate a sensitivity to synchronous proprioceptive stimulation from different body parts, i.e. the four effectors, that has a similar effect on the perceptual system as amodal stimulation across different sense modalities.

Temporal synchrony has been suggested as the "glue" that binds stimulation across the senses (Bahrick & Lickliter, 2002; Lewkowitz, 2000). In a similar manner, we imagine temporal synchrony of the movements of two or more effectors as "glue" that unifies sensations from different effectors into a coherent awareness of the body. The essential role of the limbs in early body awareness is impressively illustrated by over-represented limbs in reproduced body models with plasticine of congenitally blind infants (Kinsbourne & Lempert, 1989). The importance of temporal synchrony as well as spatial congruency for the sense of body ownership and agency has also been highlighted in studies with adults investigating the "rubber hand illusion" (Kalckert & Ehrsson, 2012). An illusory feeling of limb ownership can be induced in the absence of visual input by simultaneous proprioceptive and tactile stimulation with the participants' eyes closed (Ehrsson et al., 2005).

*4.1.4. Spatial congruence*

Spatial congruence can be experienced in two ways: First, simultaneously by spatially congruent positions - proximal or distal - of two or more effectors in a configuration. Second, temporally postponed by reliving a previous position of one or more effectors in a configuration. By returning to a previously experienced configuration, a congruent spatial state is experienced. If this occurs in conjunction with temporal synchrony, i.e. the effectors "land" simultaneously in the known congruent position, this represents a salient perceptive stimulus. Spatial congruency is an interesting point concerning the four limbs, since we have two mirror symmetrical arms and legs respectively and spatial congruency can be experienced in a gradual manner from one to four effectors.

The distribution of the configurations showed no prevailing influence of homology, bilateral- or diagonal-lateral symmetries. Interestingly, first the dimension proximal-distal proved to be a structuring factor with a gradual shift from proximal to distal in the course of development and a spurt at the age of two weeks. This shift may be a precursor of the observed change from writhing to fidgety movements (Einspieler et al., 2016). Second, the two most congruent configurations, i.e. four matched positions (0000 and 1111) were found to be significant in comparing the distributions of infants < 1 week of age and > 8 weeks. This indicates a particular importance of spatial congruence in the organization of the movement behavior since all 16 configurations are potentially available at all times. Further studies of distributions in terms of symmetries both of configurations and attractor configurations as well as their arrangement are promising, since spatial congruence plays an essential role in experiments of body ownership and self-efficacy with adults (Kalckert & Ehrsson, 2012; Ehrsson et al., 2005) and children (Rochat & Morgan, 1995).

*4.1.5. Transitions in phase space – Periods with attractor configurations and attractor-free periods*

Emergence and disappearance of preferred configurations can be understood as freezing and freeing of degrees of freedom. Freezing in sections with preferred configurations is essential for learning about the physical qualities of the body. Flexibility and plasticity in segments without attractor configurations allow the exploration of new behaviors. Thelen studied coordination modes between the limbs, which frequently fluctuated over time and described infant development as "sequences of system attractors of varying stability, evolving and dissolving over time" (p.86, Thelen & Smith, 1994, Corbetta & Thelen, 1996). This agrees with our findings of transient attractive configurations that provide pattern stability for the exploration of intrinsic limb dynamics and trajectory formation. It has long been suggested that rhythmical behavior like kicking establishes basic representations of leg movements (Piek & Gassen, 1999), functions in terms of kinesthetic calibration of the limbs (Hopkins & Prechtl, 1984) and that strong interjoint couplings provide pattern stability, within which the exploration of intrinsic limb dynamics becomes possible (Turvey & Fitzpatrick, 1993). Episodes without preferred configurations are due to self-disturbing factors that characterize dynamic systems and provide space for the development of new patterns.

Dynamically speaking, neonatal movements self-organize towards stable regions in phase space, which are sets of proprioceptive, tactile and kinesthetic inputs that act as quasi-attractors. At the same time, fluctuations in subsystems (e.g. single limbs) provide the opportunity for new combinations of sensory inputs. From the amplification of such fluctuations, recurrent passing through certain proprioceptive states occurs, and - after the cumulative process of familiarization - new attractor configurations emanate.

*4.2. Emotional processes*

Familiarization and newness usually go along with emotional or affective states that are potential components involved in transitions between periods with and devoid of attractor configurations. They are crucial to the evolving sense of body awareness in addition to perceptual and conceptual components (Berlucchi & Agliotti, 2010). "Newborns experience of the world is rich from the start. It is rich within the polarity of pleasure and pain, restfulness and agitation, approach and avoidance. They "feel" something expressing unmistakable pleasure and pains. These expressions have adaptive functions, forming crucial signals for caregivers on whom newborns rely to survive." (p.

7, Rochat, 2012). Caretakers reliably identify basic emotions expressed in the first minutes of life such as joy, pain, interest, disgust or anger. Neonates' expressions of these emotions incorporate the whole body. They become tense or spastic in certain ways and produce different sounds according to their emotional states and needs (Formby, 1967). "A rich palette of distinct affective motives underlies newborns' bodily movements and oral expressions" (p.11, Rochat, 2012).

Emergence and dissolving of attractor configurations evoke changes in perceptual patterns and this may be related to changes in emotional states. In addition to proprioceptive, tactile, and kinesthetic sensory feedback, newborns receive sensations from within their body, such as temperature, pain, muscular and visceral sensations. This interoception operates together with proprioception and exteroception, is involved in emotional expression and self-consciousness and has a neuronal correlate in the insula. The insula has been proposed to integrate all subjective feelings related to the body, including the sense of body ownership and agency, emotional experience and conscious awareness of the environment and the self (Craig, 2009).

The emergence of a new attractor configuration might be associated with feelings of interest, pleasure, and excitement of novelty that act like attractors and result in the repetition of those configurations. First, temporal synchrony and spatial congruence draw the attention and trigger feelings of interest and curiosity about the discovery of the new stimulus constellation. This is followed by the pleasure of the first reliving of newly discovered stimulus constellations. This then acts like attractors that lead to the return of these configurations.

The dissolving of recurrence patterns to certain configurations might be due to habituation to a specific configuration that is associated with feelings of indifferentism or boredom. As a result, the attractiveness of the configuration decreases, fluctuations become more prevalent, and new configurations can create interest and attraction.

*4.2.1. Hedonic topography*

In a broader sense attractor configurations can be imagined as a kind of topography of hedonic attractors that changes its shape as soon as a preferred configuration dissolves or newly emerges. The idea of a hedonic topography is aptly worded by Rochat (2012): "Neonates experience the body as an invariant locus of pleasure and pain, with a particular topography of hedonic attractors, the mouth region being the most powerful of all,[…]. Within hours after birth, in relation to this topography, infants learn and memorize sensory events that are associated with pleasure and novelty: they selectively orient to odors associated with the pleasure of feeding and they show basic discrimination of what can be expected from familiar events that unfold over time and that are situated in a space that is embodied, structured, within a body schema. But if it is legitimate to posit an a-priori "embodied" spatial and temporal organization of self-experience at birth, what might be the content of this experience aside from pleasure, pain and excitement of novelty?" (p.12, Rochat, 2012).

Attractive regions in the movement topography of the four effectors are according to the present results in the proximal-distal dimension: close to the body center and at the periphery of peripersonal space, as indicated by first, spatial and radial displacement showing the body center as a reference point of spontaneous movements. Second, the observed pattern formation emerged by partitioning the peripersonal space along the center-periphery gradient. Third, the distributions of the configurations do not show biases due to primitive reflex activity, homology or bilateral symmetry and rather a shift from configurations with proximal to distal positions with increasing age. Changing attractor configurations correspond to a changing topography of hedonic attractors within the movement field.

*4.2.2. Emotional and motivational states*

Further potential impact factors are omnipresent and changing emotional and motivational states that accompany physiological and social needs. For example "open" configurations, with more distal positions might correspond to feelings like joy and pleasure or a readiness to be in contact with caregivers, while "closed" configurations with more proximal positions might be related to a requirement of protection or orientation towards the own body. Various emotion theories assume a connection between emotional experience and movement expression. Body movements play an essential role in emotion communication (Dael, Mortillaro & Scherer, 2012) and infants coordinate emotions conveyed by body movements with affective vocalization (Zieber, Kangas, Hock & Bhatt, 2014).

*4.3. Development of motor skills*

Corbetta and Thelen (1996) analyzed effector endpoint kinematics and showed that the patterns of spontaneous arm coordination are consistent with the coordination of goal-oriented reaching movements. Accordingly, four effector configurations may be precursors of or involved in the development of goal representations. Studies with video projections of self-produced leg movements suggest a sensitivity to movement direction (Rochat & Morgan, 1995; Rochat & Striano, 2000) and neuroscientific findings in primates show that electrical stimulation of the motor cortex evoked movements toward a final configuration regardless of the starting posture (Graziano, Taylor, Moore & Cooke, 2002). Similar maps to the cortical ones can be evoked by stimulation of the spinal cord in rats and frogs (Giszter, Mussa-Ivaldi & Bizzi ,1993; Tresch & Bizzi, 2000) indicating that the spinal cord might also be sensitive to the storage of postures and the specification of trajectories.

*4.4. Conclusions*

Spontaneous movements are involved in the development of body awareness. Pattern formation displays characteristics of self-organization from which features of developing body awareness emerge that fit with concepts of the intersensory redundancy hypothesis. Turning points of hand and feet trajectories display temporal synchrony and spatial congruency and imply a sensitivity to temporal synchrony of multiple proprioceptive sensations across the body. Attractor formation, self-similarity and transitions between dynamically distinct regimes document the self-organizing quality and the intrinsic nature of the process of developing body awareness. The proximal-distal dimension proved to be crucial in comparison to reflexes, homology or bilateral-symmetry. In addition to perceptual components, emotional, motivational and cognitive aspects are postulated to be involved in pattern formation and to play a role in the development of an embodied sense of self.

5\. Acknowledgments

We thank U.M. Kahn, C. Niemitz, M. Koreuber, S. Hahn, J. Harms, M. Riesopp, A.C. Tönnissen, C. Eckrodt and the participants of the study.

Declarations of interest: None.

Funding: Humboldt Universität zu Berlin, Berliner Chancengleichheitsprogramm.

# 7. Appendix

A.1. Formulae (1-4)

Radial distance trajectories ($R_{ij}$) of the effectors $e_j$ ($j=1,2,3,4$), hands and feet, from the chest $c_i$ at each sample point $t_i$ ($i=1,2,…,n$):

(1) $R_{ij} = R_j(t_i)$ = radial distance at time $t_i$ of the *j*th effector, $e_{ij} = e_j(t_i)$, ($j = 1, 2, 3, 4$)

from the chest, $c_i = c(t_i)$

$= ((xe_{ij} - xc_i)^2 + (ye_{ij} - yc_i)^2 + (ze_{ij} - zc_i)^2)^{1/2}$

Spatial displacement ($\Delta P_{ij}$): difference in effector path distance at time point $t_{i+1}$ from time point $t_i$:

(2) $\Delta R_{ij} = \Delta R_j(t_i)$ = radial displacement of the *j*th effector during the interval from $t_{i-1}$ to $t_i$

$= R_{ij} - R_{[i-1]j}$

Radial displacement ($\Delta R_{ij}$): difference in effector distance in relation to the chest at time point $t_{i+1}$ from time point $t_i$:

(3) $\Delta P_{ij} = \Delta P_j(t_i)$ = spatial displacement of the *j*th effector along its path during the

interval from $t_{i-1}$ to $t_i$

$= ((xe_{ij} - xe_{[i-1]j})^2 + (ye_{ij} - ye_{[i-1]j})^2 + (ze_{ij} - ze_{[i-1]j})^2)^{1/2}$

Tangential velocity ($\epsilon P'_{ij}$):

(4) $\epsilon P'_{ij} = \epsilon P'_j(t_i)$ tangential velocity of the *j*th effector during the interval from $t_{i-1}$ to $t_i$

$= \Delta P_{ij}/\Delta t$, where $\Delta t$ is defined as 1/(effective sampling rate)

A.2. Databases of figures

Figure 5/6: All 14 infants/25 episodes; age [days]: M: 21.7, SD: 26.0, total time analyzed: 5h:0min:11s. Age categories: Infants < 1 week: data of 6 infants, 11 episodes, age: M: 2.1, SD: 1.0, total time analyzed: 01h:17min:02s; infants < 2 weeks: data of 7 infants, 13 episodes, age: M: 3.5, SD: 3.5, total time analyzed: 1h:46min:32s; infants > 2weeks, < 4 weeks: data of 6 infants, 6 episodes, age: M: 23.2, SD: 3.9, total time analyzed: 1h:16min:31s; infants > 4 weeks: data of 5 infants, 6 episodes, age: M: 59.8, SD: 24.2, total time analyzed: 1h:57min:09s; infants > 8 weeks: data of 3 infants, 3 episodes, age: M: 79.5, SD: 11.6, total time analyzed: 46min:39s.

Figure 9: All 14 infants/25 episodes; age [days]: M: 21.7, SD: 26.0, total time analyzed: 5h:0min:11s. 8745 time windows (1197 with 300dp = 60s, 7445 with 750dp = 30s for group 2 and accordingly 15s for group 3 infants); infants < 2 weeks: data of 7 infants, 13 episodes, age: M: 3.5, SD: 3.5, total time analyzed: 1h:46min:32s; 1115 time windows with 300dp = 60s; infants > 2 weeks: 8 infants, 12 episodes, age: M: 41.5, SD: 25.3, total time analyzed: 3h:13min:40s, 7527 time windows (82 with 300dp = 60s, 7445 with 750dp = 30s for group2 and accordingly 15s for group 3 infants).